\documentclass[%
 aip,
cp,  
 amsmath,amssymb,
 reprint,%
]{revtex4-2}

\usepackage{graphicx}
\usepackage{dcolumn}
\usepackage{bm}

\usepackage[utf8]{inputenc}
\usepackage[T1]{fontenc}
\usepackage{mathptmx} 

\begin{document}

\title{Inverse design of ultrathin metamaterial absorber}
\author{Eunbi Jang\dag} 
 \affiliation{Department of Artificial Intelligence Semiconductor Engineering, Hanyang University, Seoul, 04763, South Korea}
\author{Junghee Cho\dag}%
 \affiliation{Department of Electronic Engineering, Hanyang University, Seoul, 04763, South Korea}

\author{Chanik Kang}
 \email[Corresponding author:]{chanik@hanyang.ac.kr}
\affiliation{%
Department of Artificial Intelligence, Hanyang University, Seoul, 04763, South Korea}%

\author{Haejun Chung}
 \email[Corresponding author:]{haejun@hanyang.ac.kr}
\affiliation{Department of Artificial Intelligence Semiconductor Engineering, Hanyang University, Seoul, 04763, South Korea}
\affiliation{Department of Electronic Engineering, Hanyang University, Seoul, 04763, South Korea}
\affiliation{%
Department of Artificial Intelligence, Hanyang University, Seoul, 04763, South Korea}%

\begin{abstract}
Electromagnetic absorbers combining ultrathin profiles with robust absorptivity across wide incidence angles are essential for applications such as stealth applications, wireless communications, and quantum computing. Traditional designs, including Salisbury screens, typically require thicknesses of at least a quarter-wavelength ($\lambda/4$), restricting their use in compact systems. While metamaterial absorbers (MMAs) offer reduced thicknesses, their absorptivity generally decreases under oblique incidence conditions. Here, we introduce an adjoint optimization-based inverse design method that merges the ultrathin advantage of MMAs with the angle-insensitive characteristics of Salisbury screens. By leveraging the computational efficiency of the adjoint method, we systematically optimize absorber structures as thin as $\lambda/20$. The optimized structures achieve absorption exceeding 90\% at the target frequency (7.5 GHz) and demonstrate robust performance under oblique incidence, maintaining over 90\% absorption up to 50$^\circ$, approximately 80\% at 60$^\circ$, and around 70\% at 70$^\circ$. Comparative analysis against particle swarm optimization further highlights the superior efficiency of the adjoint method, reducing the computational effort by approximately 98\%. This inverse design framework thus provides substantial improvements in both performance and computational cost, offering a promising solution for advanced electromagnetic absorber design.

\noindent$^\dag$These authors contributed equally to this work.

\end{abstract}

\maketitle
 
\section{1. Introduction}
Metamaterials are artificially engineered subwavelength structures designed to exhibit electromagnetic properties unachievable in naturally occurring bulk materials~\cite{smith2004metamaterials, chung2020high}. Unlike conventional materials, whose electromagnetic characteristics are dictated primarily by atomic composition, metamaterials derive their unique properties from deliberately designed subwavelength architectures, often arranged periodically. By controlling their effective permittivity ($\varepsilon$) and permeability ($\mu$)—parameters which govern electromagnetic wave propagation, metamaterials enable extraordinary manipulation of wave behaviors. Metamaterials and their two-dimensional (2D) counterparts, metasurfaces, have demonstrated superior performance in diverse applications, including metalenses~\cite{lin2019topology, phan2019high, seo2024deep}, wavefront shaping~\cite{kang2025adjoint}, electromagnetic sensing~\cite{wan2024information, wang2021wearable}, wave absorption~\cite{lee2024broadband, akselrod2015large, ran2022optically}, and light extraction~\cite{vaskin2019light, bae2022enhanced, choi2022simple}, and many others~\cite{gieseler2024chiral, xia2024nonlinear, latifpour2024hyperspectral, carter2024flat, li2023metasurfaces, shao2025dual}.

Electromagnetic perfect absorption has gain significant interest due to its broad industrial applications, including electromagnetic compatibility~\cite{landy2008perfect, watts2012metamaterial}, electromagnetic shielding~\cite{fante1988reflection, kim2017selective}, and the cloaking of unmanned aerial vehicles (UAVs)~\cite{dewangan2023broadband}. These applications typically demand ultrathin absorbing layers to achieve high performance while maintaining a compact form. The Salisbury Screen, introduced in the 1950s, demonstrated over 90\% absorption within a narrow frequency band using a $\lambda/4$-thick structure~\cite{salisbury1952absorbent}. This concept was later extended to Jaumann absorbers~\cite{du1994design}, which consist of multiple layers of Salisbury Screens to achieve broadband absorption. However, both designs require a thickness equal to or exceeding $\lambda/4$ to maintain their intended performance, posing a critical limitation for gigahertz-frequency applications and UAV integration, where minimizing structural bulk is essential~\cite{rozanov2000ultimate, kang2023multispectral}.
In contrast, metamaterial absorbers (MMAs) offer the potential for high absorption efficiency with ultrathin layers, making them promising candidates for alternative compact absorbers~\cite{lyu2021transparent, deng2021ultra, song2023inverse}. However, these designs inherently suffer from performance degradation due to their sensitivity to both incidence angle and wavelength, a consequence of their design methodology. The periodic arrangement of meta-atoms simplifies the design process, as an optimized single meta-atom can be replicated across a large area, eliminating the need to simulate or optimize the entire structure. However, in the case of oblique incidence or longer wavelengths, the interactions between neighboring meta-atoms lead to significant performance deterioration, limiting the effectiveness of metamaterial absorbers in broadband and wide-angle applications.

\begin{figure}[!htp]
\includegraphics[width=\textwidth]{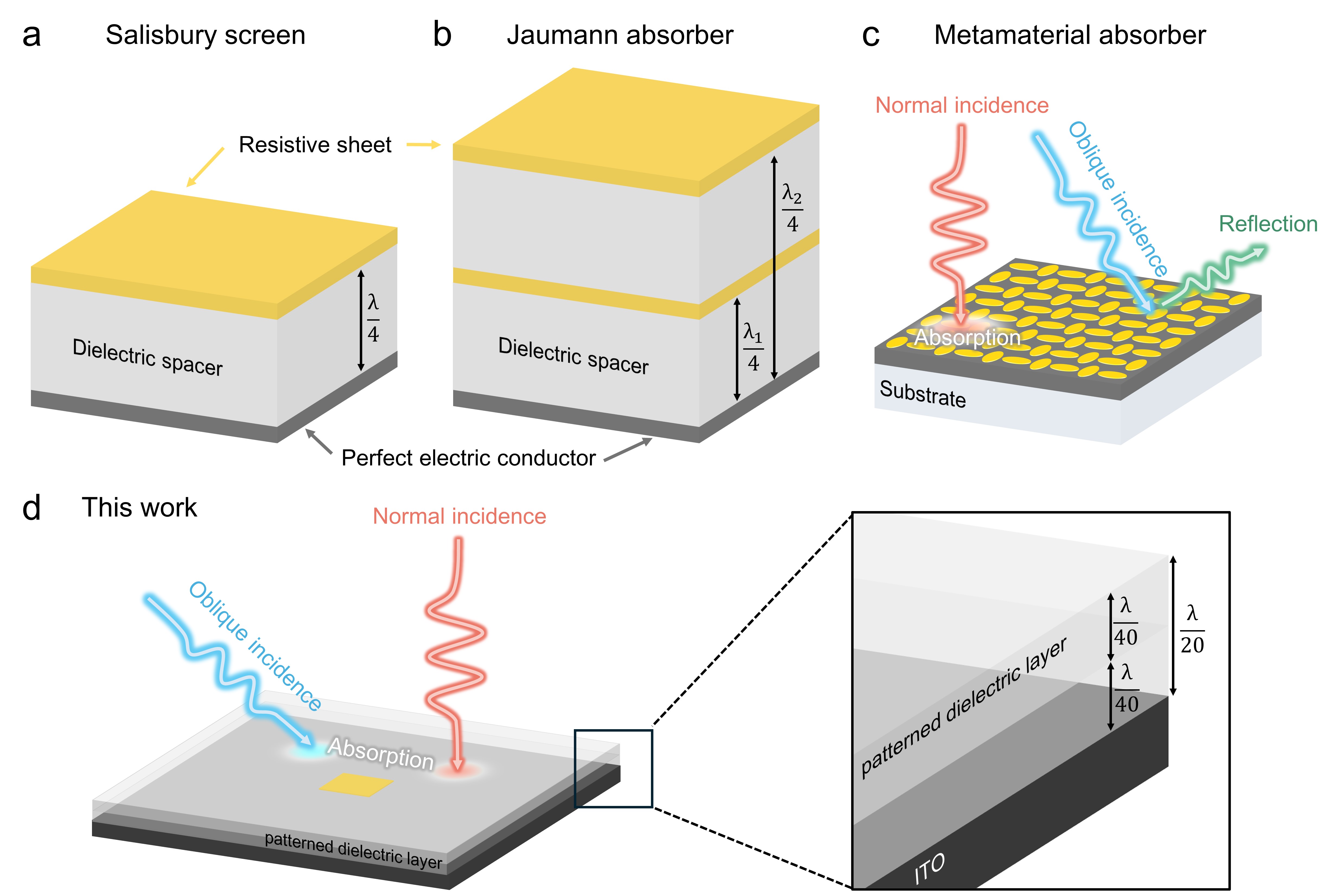}
\caption{(a) Schematic illustration of a classical Salisbury screen, which relies on a quarter-wavelength ($\lambda/4$) dielectric spacer and a resistive sheet over a perfect electric conductor (PEC) to achieve narrowband absorption~\cite{salisbury1952absorbent}. (b) The Jaumann absorber, formed by stacking multiple Salisbury screens, broadens the absorption bandwidth at the expense of increased overall thickness~\cite{du1994design}. (c) A metamaterial absorber, which uses subwavelength resonators to significantly reduce thickness but often suffers from narrow bandwidth and sensitivity to oblique incidence. (d) The approach introduced in this work, merging the ultrathin advantage of metamaterial absorbers with the wide-angle robustness of Salisbury screens via an adjoint-based inverse design method, thereby enabling a $\lambda/20$-thick absorber with high absorption over a broad range of incidence angles.}
\label{fig1}
\end{figure} 

In the electromagnetics area, inverse design refers to a computational method in which the desired figure of merit (FoM) is defined first and an optimization algorithm is employed to determine the optimal structure that achieves the specified performance~\cite{kim2025freeform}. Specifically, we utilize adjoint optimization, a large-scale computational technique that efficiently computes gradients in high-dimensional design problems. This method is particularly advantageous in electromagnetics, enabling gradient-based optimization with minimal computational overhead~\cite{mansouree2021large, kang2024large}.

In this work, we develop an adjoint method-based inverse design framework for combining the advantages of the metamaterial absorbers and Salisbury Screen, where the former utilizes an ultrathin layer, and the latter offers robustness over wide incidence angles. 
We address the design problem of translation-symmetric metamaterial absorbers (MMAs) integrated with a $\lambda/20$-thick Salisbury Screen~\cite{salisbury1952absorbent}. Next, we analyze the angle-dependent characteristics of the optimized MMAs to assess their performance under varying incidence conditions. Subsequently, we investigate fully three-dimensional (3D) MMA designs, employing particle swarm optimization (PSO) and adjoint optimization for comparative analysis. PSO is a widely used global optimization technique in electromagnetics that has demonstrated strong performance when optimizing a limited number of parameters~\cite{kennedy1995particle}. However, it faces significant challenges in handling complex, high-dimensional design problems~\cite{rini2011particle}.

Finally, we validate that our adjoint method-based framework achieves superior device performance while significantly reducing computational cost compared to PSO. We believe that this framework can be extended to a broad range of MMA design problems as well as other electromagnetic applications, offering an efficient approach to optimizing complex photonic and electromagnetic structures.
Our optimized ultrathin absorber structure, designed at a reduced thickness of $\lambda/20$, achieved absorption efficiencies exceeding 90\% at the target frequency of 7.5~GHz, demonstrating robust performance at oblique incidences (over 90\% up to 50$^\circ$, approximately 80\% at 60$^\circ$, around 70\% at 70$^\circ$) while reducing the computational cost by approximately 98\% compared to PSO.

\section{2. Methods}

Inverse design approaches for electromagnetic absorbers typically rely on forward methods~\cite{landy2008perfect}, evolutionary algorithms~\cite{gold2024gaga, chung2020tunable}. The computational cost of forward methods, such as parameter sweeps~\cite{wu2016low, mahmud2020wide}, increases rapidly with the number of design parameters, becoming impractical for large parameter spaces~\cite{kang2024large}. On the other hand, evolutionary algorithms, including genetic algorithms (GA)~\cite{holland1992genetic} and PSO~\cite{kennedy1995particle}, provide robustness in solving complex, non-convex design space problems but require a substantial number of evaluations, resulting in significant computational overhead.

\begin{figure}[!htp]
\includegraphics[width=\textwidth]{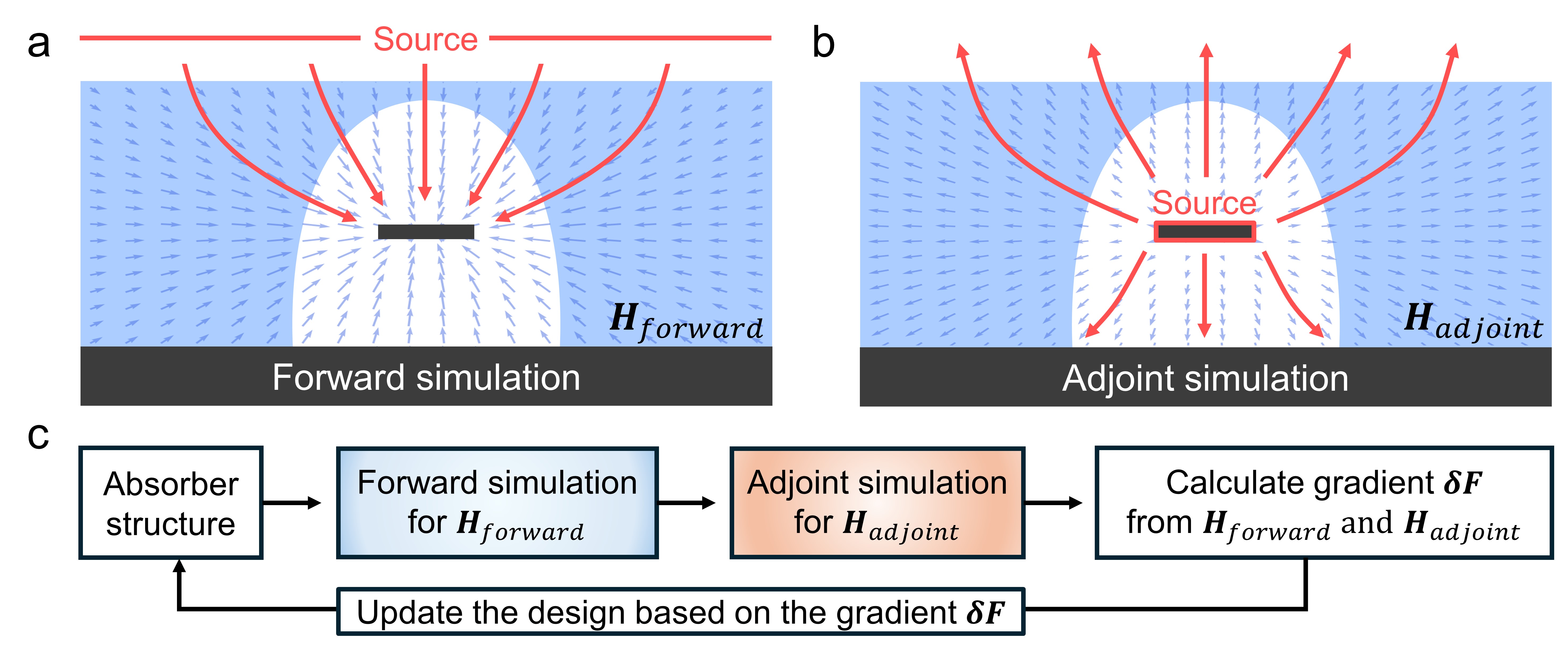}
\caption{Schematics of the forward and adjoint simulations and the iterative optimization procedure. (a) Forward simulation configuration, in which a planewave source illuminates the design region. The objective is to maximize the magnetic field intensity at the central target area, thereby enhancing electromagnetic absorption. The blue arrows represent the magnetic field distribution vectors within the target structure. (b) Adjoint simulation configuration used in the adjoint method-based inverse design. Contrary to the forward simulation, the adjoint simulation employs dipole sources positioned at the target region, propagating fields outward through the design domain to calculate sensitivity gradients efficiently. (c) Schematic illustration of the iterative optimization process based on adjoint gradient calculations, where the structure progressively evolves toward optimal electromagnetic absorption performance.}
\label{fig2}
\end{figure} 

To overcome the limitations of conventional design methods, the adjoint method has emerged as a powerful alternative for the inverse design of electromagnetic devices~\cite{Miller:EECS-2012-115, ji2024efficient}. As demonstrated in Fig.~\ref{fig2}, this approach leverages Lorentz reciprocity and Born approximation to compute the gradient of a specified objective function with respect to all design parameters simultaneously, comprising a forward simulation and an adjoint simulation. This enables solving large-scale complex optimization problems, even having billions of design parameters~\cite{aage2017giga}. In contrast to the brute-force method, which necessitates separate simulations for each design variable, the adjoint method obtains the entire gradient information with only two simulations (forward and adjoint simulations), dramatically reducing computational cost~\cite{Miller:EECS-2012-115}. 

For electromagnetic absorber design problems, we define a performance metric (figure-of-merit, FoM) over a target volume \(V_x\) (e.g., ITO regime), with the design domain denoted by \(V_{x'}\) (e.g., Si and SiO$_{\text{2}}$ regime). Typically, dielectric absorbers employ a conventional figure‐of‐merit expressed as $\frac{\int |E(x)|^2\, {dV_x}}{P_{\text{inc}}}$, which is intuitive since $\int \varepsilon'' |E(x)|^2 \,{dV_x}$ corresponds to absorbed power~\cite{miller2016fundamental}. However, our absorber consists of a monolayer metallic film where electric field discontinuities at interfaces complicate analysis due to its high conductivity contrast. Therefore, we define an alternative FoM using magnetic fields and dipoles to enhance absorption in the metallic film:
\[
F = \int_{V_x} f(E(x), H(x))\, d{V_x},
\]
where \(f(E(x), H(x))\) represents the local performance metric (e.g., field intensity or absorbed power density), and the design objective is to maximize the integrated response over the target domain \(V_x\), where the figure-of-merit is defined and evaluated, with coordinates \(x\).

When the design is perturbed by the inclusion of a small optical feature, the electromagnetic fields change by tiny amounts \(\delta E\) and \(\delta H\). Therefore, we can employ linearization of the merit function with respect to these perturbations yields
\[
\delta F 
= \int_{V_x}
\Biggl[
\frac{\partial f}{\partial \overline{E}(x)} \cdot \delta \overline{E}(x)
+ \frac{\partial f}{\partial E(x)} \cdot \delta E(x)
+ \frac{\partial f}{\partial \overline{H}(x)} \cdot \delta \overline{H}(x)
+ \frac{\partial f}{\partial H(x)} \cdot \delta H(x)
\Biggr]
\, dV_x.
\]
Since \(F\) is real value while the fields are complex-valued here, we combine the corresponding complex-conjugate terms to obtain
\[
\delta F = 2\,\operatorname{Re}\left\{ \int_{V_x} \left[ \frac{\partial f}{\partial E}(x) \cdot \delta E(x) + \frac{\partial f}{\partial H}(x) \cdot \delta H(x) \right] {dV_x} \right\}.
\]
We assume that \(\delta E\) and \(\delta H\) are sufficiently small such that higher-order terms can be neglected~\cite{Miller:EECS-2012-115}.

In the design of our electromagnetic absorber, the metallic region (denoted by \(x\)) exhibits pronounced discontinuities in the electric field due to abrupt material property changes at the interface due to the high conductivity value of the ITO. Consequently, the electric field perturbation \(\delta E\) is cumbersome and less reliable for sensitivity analysis. To circumvent these difficulties, the magnetic field response is more continuous and amenable to modeling. Therefore, we intentionally omit the electric field contribution and retain only the magnetic field term, yielding the simplified sensitivity expression
\[
\delta F \approx 2\,\operatorname{Re}\left\{ \int_{V_{x}} [\frac{\partial f}{\partial H}(x) \cdot \delta H(x)]\, {dV_{x}} \right\}.
\]

To quantify the influence of magnetic field perturbations on the performance metric, we introduce an adjoint magnetic field, denoted by \(H_{\text{adj}}(x')\). Exploiting the symmetry of the Green's function, this adjoint field, which encapsulates the sensitivity information associated with magnetic dipole sources, is defined as
\[
H_{\text{adj}}(x') = \int_{V_x} G(x', x)\,\frac{\partial f}{\partial H}(x)\, {dV_x},
\]
where \(G(x', x)\) is the magnetic Green's function mapping the response of a unit magnetic dipole at \(x\) (within the metallic absorber) to the magnetic field at a point \(x'\) in the design region, and \(\frac{\partial f}{\partial H(x)}\) represents the local sensitivity of \(f\) concerning \(H\).

For our grid‐based (rectangular) design in FDTD, the inclusion is represented by a cell of volume \(\Delta V\). For a cell located at \(x'\) in the design domain, the induced dipole moment is approximated as
\[
p_{\text{ind}}(x') \approx \Delta V\, (\varepsilon_2 - \varepsilon_1)\, E_{\text{old}}(x'),
\]
where \(E_{\text{old}}(x')\) denotes the unperturbed (or baseline) electric field in the design domain prior to the introduction of the dielectric perturbation. This reference field is essential for quantifying the first‐order change in \(F\) via the induced polarization. Accordingly, the polarization density (dipole moment per unit volume) is given by
\[
P_{\text{ind}}(x') = \frac{p_{\text{ind}}(x')}{\Delta V} \approx (\varepsilon_2 - \varepsilon_1)\, E_{\text{old}}(x').
\]

For a small cubic cell of volume \(\Delta V\) in the design domain, one would characterize a local change in permittivity by introducing an effective polarizability ($ \alpha_{\text{cube}}$)
\[
P_{\text{ind}}(x') = \alpha_{\text{cube}}\, E_{\text{old}}(x').
\]

To calculate the effective polarizability of the cubic inclusion, one would use the Clausius-Mossotti relation for certain geometries (e.g., sphere, cylinder, cubic, etc). Following the Clausius–Mossotti approximation for spherical inclusions~\cite{jackson1998classical, sihvola1999electromagnetic, Miller:EECS-2012-115}, the effective polarizability for a cubic inclusion can be expressed by setting the depolarization factor \(L \approx 1/3\) corresponding to cubic symmetry:
\[\alpha_{\text{cube}} \approx \Delta V \varepsilon_1 \frac{\varepsilon_2 - \varepsilon_1}{\varepsilon_1 + L(\varepsilon_2 - \varepsilon_1)},
\]
where \(\varepsilon_1\) and \(\varepsilon_2\) represent the permittivities of the background and the inclusion, respectively.

Substituting this expression into the first-order variation of the figure-of-merit and utilizing the adjoint magnetic field, we obtain
\[
\delta F = 2\,\operatorname{Re}\left\{ \int_{V_{x}} \, dV_x \int_{V_{x'}}\, d{V_{x'}}  \left[P_{\text{ind}}(x')\, \cdot G(x, x')\,\frac{\partial f}{\partial H}(x) \, \right]
 \right\}.
\]
The notation $\int_{V_{x}}$\(dV_{x}\) and $\int_{V_{x'}}$\(dV_{x'}\) explicitly indicate volume integrals carried out over these distinct regions, respectively. Thus, the expression for \(\delta F\) to
\[
\delta F \approx 2\,\operatorname{Re}\left\{\int_{V_{x}} \, dV_x\int_{V_{x'}} P_{\text{ind}}(x') \cdot H_{\text{adj}}(x')\, dV_{x'}\right\}.
\]
This final expression indicates that the variation in the performance metric is directly proportional to the effective polarizability of the cell, the unperturbed electric field, and the adjoint magnetic field. Evaluating this sensitivity over the entire design domain provides the optimal modifications required to maximize \(F\), thereby establishing an efficient adjoint-based inverse design method.

Based on the sensitivity expression derived above, iterative adjoint-based optimization can be implemented to refine the metamaterial-based absorber design systematically. First, the forward simulation provides the electromagnetic fields, and subsequently, the adjoint simulation generates sensitivity fields. These sensitivity fields quantify how local permittivity changes impact the performance metric \(F\)~\cite{kang2024adjoint}. Each iteration updates the permittivity distribution in a gradient-ascent fashion:
\[
\phi^{(n+1)}(x') = \phi^{(n)}(x') + \eta\,\frac{\delta F(x')}{\delta \phi(x')},
\]

where \(\phi^{(n)}(x')\) represents the permittivity distribution at the \(n\)-th iteration, \(\eta\) is the learning rate, and \(\frac{\delta F(x')}{\delta \phi(x')}\) is the gradient of the figure-of-merit with respect to the permittivity at position \(x'\).

The adjoint optimization typically results in a grayscale permittivity profile without any external binarization term. Therefore, a binarization step is subsequently employed to enforce discrete, manufacturable permittivity values~\cite{bae2023inverse}. This iterative adjoint optimization and binarization process continues until convergence, yielding an optimized monolayer metallic absorber structure with high electromagnetic performance.

\section{3. Results \& Discussion}
\subsection{3.1. Two-dimensional absorber optimization}
We first demonstrate the adjoint method-based inverse design technique by applying it to a 2D metamaterial absorber. The optimization is performed under normal incidence to maximize electromagnetic absorption at the target frequency. The optimized structure is then evaluated using finite-difference time-domain (FDTD) simulations to assess its absorption performance. We further examine the design under oblique incidence to test its angular robustness. This evaluation confirms that the proposed framework achieves high absorption efficiency and highlights its practical applicability.

\begin{figure}[!htp]
\includegraphics[width=\textwidth]{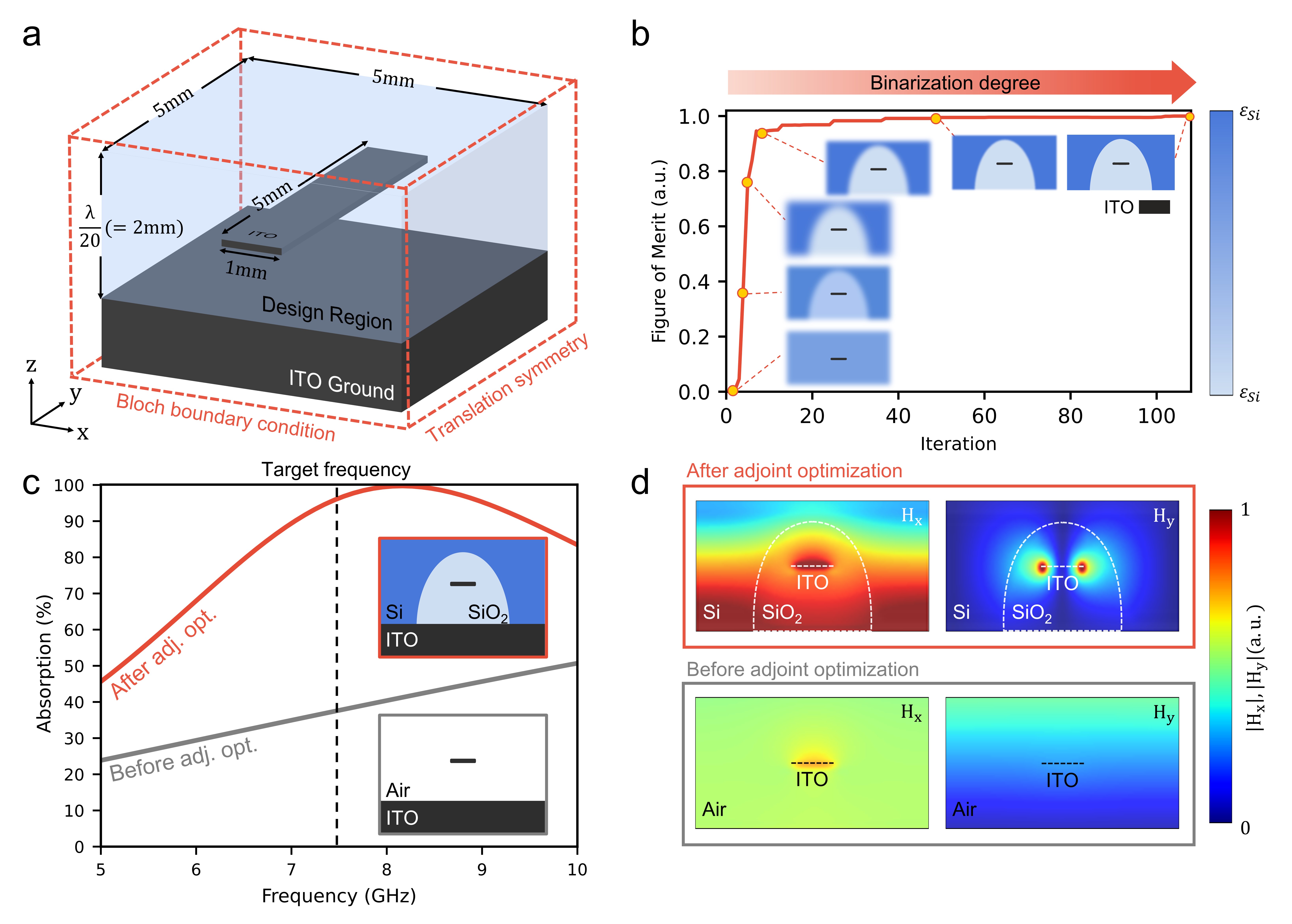}
\caption{Two-dimensional metamaterial absorber design.
(a) The 2D simulation domain measures 5\,mm in width and 2\,mm in height, providing a total thickness of $\lambda/20$. The region surrounding the ITO patch is defined as the optimization domain.
(b) Evolution of the FoM and the binarization process during optimization. Initially, the structure is assigned an intermediate permittivity; as the optimization progresses, it gradually transitions to either silicon (Si) or silicon dioxide (SiO$_2$).
(c) Absorption spectra before (gray) and after (red) optimization, showing a marked increase at the target frequency of 7.5\,GHz.
(d) Electromagnetic field distributions before (bottom) and after (top) optimization. In the optimized structure, the magnetic field $H$ is strongly concentrated around the ITO patch, resulting in high absorption performance.
}
\label{fig3}
\end{figure} 

\subsubsection{3.1.1. Optimization under normal incidence}
We apply the adjoint-based inverse design framework to a 2D metamaterial absorber operating at 7.5~GHz—a frequency band relevant to radar stealth and RF applications~\cite{zeng2020electromagnetic}. The absorber structure consists of an ultrathin dielectric layer with a thickness of $\lambda/20$ (2~mm), and the optimization is performed to maximize absorption under normal incidence. We perform simulations using the finite-difference time-domain (FDTD) method with a grid spacing of 0.125~mm. The simulation domain spans 5~mm in the $x$-direction with periodic boundary conditions, and 2~mm along the $z$-axis with perfectly matched layer (PML) boundaries. A Gaussian beam source centered at 7.5~GHz is used to excite the structure~\cite{oskooi2010meep}. This setup enables a detailed examination of the absorber’s resonant response and the spatial distribution of the electromagnetic fields.

As shown in Fig.~\ref{fig3}(a), the basic configuration of the absorber comprises a patch—which is mainly responsible for electromagnetic absorption, and a ground (GND) layer, separated by a 1 mm gap. Before applying the optimization procedure, we set the baseline result by simulating the patch and the air-filled GND layer. The absorptive metallic patch is assumed to be Indium Tin Oxide (ITO) (conductivity=$0.5 \times 10^{6}$  S/m, relative permittivity=$0.5 \times 10^{6}$ at 7.5~GHz)~\cite{alwan2015indium}, where it is known for its high electrical conductivity at microwave band. However, the ITO patch is extremely thin, with a thickness of only 25~nm, which corresponds to the $6.25 \times 10^{-7} \times \lambda$ at 7.5~GHz. Since the design regime has $5.00 \times 10^{-2}\lambda$ thickness, we face multi-scale simulation and optimization problems.

For example, modeling the ITO layer at its actual 25~nm thickness would require a simulation grid spacing of 25 nm. When the simulation domain is confined to the essential field propagation region and the absorber structure measures 20 mm by 5 mm, the resulting total of approximately 160 billion pixels renders the simulation computationally infeasible.

In computational electromagnetics, graphene and other 2D materials are often modeled with a delta function conductivity in Maxwell's equations because their physical thickness is negligible relative to the wavelength~\cite{castro2009electronic}. These materials typically exhibit anisotropic conductivity, which generates currents parallel to the surface in response to the surface-parallel components of the electric field. In a discretized simulation environment like FDTD, this idealization is approximated by implementing an anisotropic volume conductivity (or another form of polarizable dispersive material) whose effective thickness scales as a single grid spacing, with its amplitude scaled inversely proportionally to the grid spacing. For instance, a one-pixel-thick ITO can be modeled using a single grid spacing thick surface while explicitly dividing the conductivity value by the grid spacing~\cite{eriksson2007efficient, bendsoe2013topology}. This method provides an efficient and accurate means to incorporate the electromagnetic behavior of ultrathin ITO layers in the simulation.

Accordingly, we modeled the ITO layer as a monolayer with a thickness of 125~$\mu$m and set the simulation grid spacing to 125~$\mu$m. To accurately replicate the electrical properties of the actual 25~nm-thick ITO, we scaled the conductivity and permittivity of the 125~$\mu$m layer to approximately $1/5000$ of their original values, reflecting the thickness ratio between 25~nm and 125~$\mu$m. We validated this approximation by comparing two simulation approaches: one approach modeled a 25~nm-thick ITO using the Transfer Matrix Method (TMM), while the other simulated the 125~$\mu$m-thick, conductivity-tuned ITO using FDTD. The reflection, transmission, and absorption spectra from both methods closely agreed, exhibiting only a marginal numerical error (approximately $0.23\%$). This modeling strategy reduced the number of simulation pixels by $1/5000^2$, significantly optimizing the required computational resources.

Although the optimization design and simulation are conducted on a single periodic unit cell, Bloch boundary conditions enable the simulation to represent an environment where the structure is infinitely repeated along the $x$-axis. For simplicity, the horizontal dimension is set to 5~mm. While further investigation into the optimal periodicity and patch dimensions might enhance performance, we maintain a fixed 5-mm periodicity and patch size to demonstrate the robustness of our proposed method across various optimization setups. Moreover, given the translation symmetry along the $y$-axis, the optimization process does not consider any variations in that direction.

The optimization procedure demonstrates rapid improvement in the FoM, highlighting the adjoint method as an efficient inverse design strategy. Initial design domains begin with grayscale permittivity distributions, as illustrated in the inset at the lower left of Fig.~\ref{fig3}(b). Adjoint derivatives, $\frac{\delta \mathrm{FoM}}{\delta \epsilon(x)}$, quantify sensitivity to local permittivity variations and actively guide early optimization stages. Notably, the FoM rapidly increases within the first ten iterations, underscoring the computational efficiency and accuracy of the adjoint-based framework. A binarization factor~\cite{pan2025inverse} subsequently binarizes the permittivity distribution into Si and SiO$_2$. The optimization process converges fully within approximately 50 iterations, retaining negligible differences between pre- and post-binarization configurations. Thus, the proposed approach significantly reduces computational resources compared to conventional methods, enabling rapid convergence.

As demonstrated in Fig.~\ref{fig3}(c), the absorption performance dramatically improves at the target frequency of 7.5 GHz after optimization, increasing from an initial 35\% to approximately 95\%. The baseline structure, composed only of a 1-mm air gap between the conductive patch and ground (GND) layer, typically achieves peak absorption far above the desired frequency (15--20 GHz) based on the Salisbury screen principle. Consequently, at 7.5 GHz, absorption remains insufficient. In contrast, the optimized absorber incorporates an ultrathin dielectric layer, achieving a total thickness of $\lambda/20$. Despite its reduced thickness, careful permittivity arrangement successfully emulates a refractive mechanism equivalent to a classical Salisbury screen with a much thicker dielectric ($\lambda/4$). This ultrathin, hybrid Salisbury-metamaterial structure significantly expands potential applications requiring compact and angle-insensitive absorbers.

Electromagnetic field distributions clearly illustrate in Fig.~\ref{fig3}(d), performance enhancements achieved by the optimized absorber. Initially, the baseline design shows weak magnetic field components ($H_x$ and $H_y$) near the conductive patch. Weak fields result in insufficient surface plasmon resonance (SPR), limiting electromagnetic absorption efficiency. However, the optimized structure substantially strengthens magnetic field components around the conductive patch. Notably, a strong concentration of $H_y$ fields emerges at the edges of the indium tin oxide (ITO) patch, revealing a previously unreported resonance mode enhancing absorptivity. These intensified fields induce pronounced SPR phenomena, converting electromagnetic energy into heat more effectively and thereby significantly reducing reflection. This newly identified field-concentration mechanism highlights an important physical insight that generalized dielectric arrangements can notably amplify absorption performance. Consequently, this framework not only optimizes absorber designs computationally but also provides a powerful tool to explore novel electromagnetic phenomena applicable to various materials and structural configurations.

Consequently, our optimized 2D metamaterial absorber achieves exceptional absorption at the target frequency of 7.5 GHz, even with a reduced dielectric thickness of only $\lambda/20$.

\subsubsection{3.1.2. Performance evaluation under oblique incidence}

\begin{figure}[!htp]
\includegraphics[width=\textwidth]{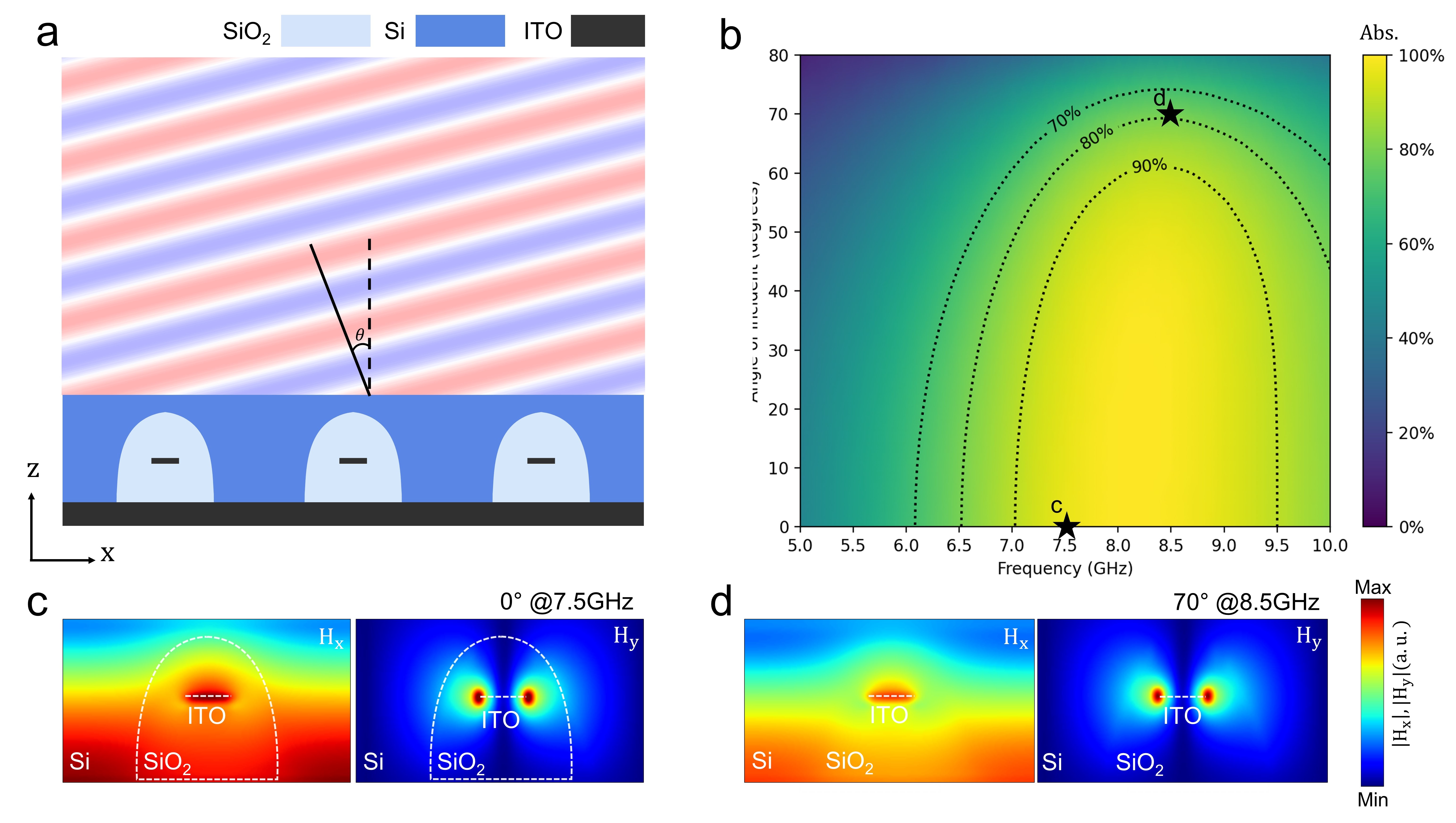}
\caption{
Oblique incidence performance of the 2D optimized absorber. (a) Cross-sectional schematic illustrating the oblique incidence, depicting a planewave striking the ultrathin absorber structure.
(b) Absorption performance map presented as a function of frequency (5--10 GHz) and incidence angle (0--80$^\circ$). The optimized absorber maintains exceptional robustness, providing over 90\% absorption efficiency within a broad frequency range of approximately 7.0--9.6 GHz, even at angles up to 50$^\circ$. Significantly, absorption remains substantial at higher incidence angles, demonstrating approximately 80\% at 60$^\circ$ and around 70\% even at 70$^\circ$.
(c,d) Electromagnetic field distribution comparisons highlighting consistent and robust field localization under different incident conditions:
(c) Normal incidence (0$^\circ$) at the target frequency of 7.5 GHz, showing intense magnetic field $H_y$ concentration primarily near the edges of the ITO patch;
(d) Oblique incidence at a high angle of 70$^\circ$ at 8.5 GHz, $H_y$ still showing substantial field localization near the patch edges despite the increased angle.
}
\label{fig4}
\end{figure} 

In practical applications of electromagnetic absorbers, such as radar stealth technology, RF systems, and wireless power transfer, the absorber structures often encounter waves that are not normal to the surface. To ensure effective operation under such conditions, the absorber must maintain high performance over a wide range of incident angles.\\
To assess the angular robustness of our optimized 2D absorber (described in Section~2.2.1.), we performed simulations using a plane wave source with incident angles varying from $0^\circ$ to $80^\circ$. As illustrated in Fig.~\ref{fig4}(a), the schematic clearly depicts an oblique incidence scenario where a plane wave strikes the ultrathin absorber. The corresponding absorption performance map in Fig.~\ref{fig4}(b) shows that at the target frequency, the absorber maintains an absorption efficiency exceeding 90\% for incident angles between $0^\circ$ and $50^\circ$, while the efficiency decreases gradually to approximately 80\% at $63^\circ$ and around 70\% at $70^\circ$. 

Furthermore, the electromagnetic field distribution comparisons in Fig.~\ref{fig4}(c) and \ref{fig4}(d) reveal that under normal incidence (0$^\circ$) the magnetic field $H_y$ is intensely concentrated near the edges of the ITO patch, and even at a high oblique angle (70$^\circ$) the field localization remains substantial.
Based on these comprehensive observations, we extended the analysis to a 3D configuration to further optimize the design and evaluate its performance under more realistic conditions.

\subsection{3.2. Three-dimensional absorber optimization}
In this section, we extend the results from the 2D design to a 3D structure that reflects real-world application environments, utilizing the computationally efficient adjoint method-based optimization. This demonstrates the feasibility of designing ultrathin metasurface metamaterial absorbers with excellent performance, even in complex environments with a large number of design variables.

The design aims to achieve high absorption performance in a specific frequency range. We performed the simulations using Meep, an open-source FDTD tool. The simulation grid spacing is 0.125 mm, and the size of the x-y plane is 5 mm by 5 mm, with periodic boundary conditions applied in both the x and y directions. The thickness along the z-axis is set to 2 mm ($\lambda/20$ at 7.5~GHz), with PML boundary conditions applied in the z-direction. The incident plane wave is excited using a Gaussian source with a central frequency of 7.5~GHz, at a distance of $0.3\lambda$ from the design region along the z-axis. Since the number of design parameters exceeds 30,000, this can be considered a complex optimization problem. The FoM for the inverse design of the combined Salisbury screen and metamaterial absorber is defined as follows.

\begin{figure}[!htp]
\includegraphics[width=\textwidth]{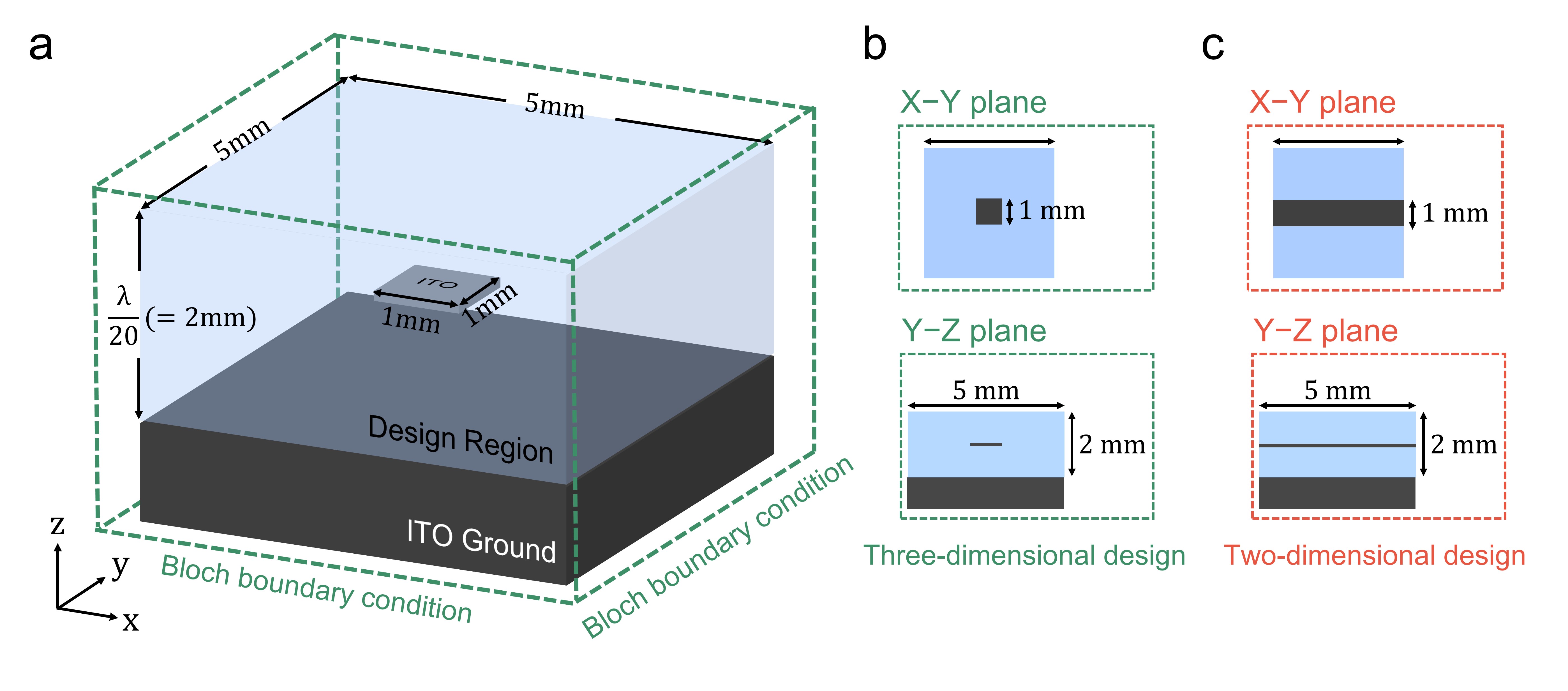}
\caption{
 Schematic of the optimized Three-demensional metamaterial absorber design.
(a) The 3D simulation domain measures 5\,mm $\times$ 5\,mm in the $xy$-plane and 2\,mm in the $z$-direction (providing a total thickness of $\lambda/20$ at 7.5~GHz). Bloch boundary conditions are applied along the $x$- and $y$-axes, while a perfectly matched layer (PML) is employed along the $z$-axis. A square ITO patch is placed at the center, and the region surrounding the patch is defined as the optimization domain.
(b) Cross-sectional views of the 3D structure in (a) along the $xy$- and $yz$-planes. 
(c) Cross-sectional views of the previously introduced 2D structure in the $xy$- and $yz$-planes.
}
\label{fig5}
\end{figure} 

The simulation structures proposed in this study can be broadly classified into a single unified design, as shown in Fig.~\ref{fig5}. In Fig.~\ref{fig5}(a), the optimized 3D absorber is presented. In this configuration, Bloch boundary conditions are applied along the $x$- and $y$-axes, while PML is employed along the $z$-axis. The design region measures 5\,mm $\times$ 5\,mm in the $xy$-plane and has a thickness of $\lambda/20$, with a square ITO patch placed at its center. This 3D configuration, with the design region filled with air, serves as the benchmark for the 3D optimization design. In contrast to the 2D structure, which assumes invariance along one axis, the 3D design accommodates a significantly larger number of degrees of freedom, thereby reflecting more realistic absorber geometries.
The simulation environment (e.g., incident wave, flux monitors, grid spacing, simulation domain, design variables, and material properties) is maintained identical to that of the 2D design. Fig.~\ref{fig5}(b) shows cross-sectional views of the 3D absorber along the $xy$- and $yz$-planes, while Fig.~\ref{fig5}(c) presents cross-sectional views of the previously described 2D absorber.
By comparing these cross-sections, one can observe how the 3D design extends the optimization domain along both $x$ and $y$ axes, enabling complex field interactions that cannot be captured in the 2D setup.

We determine the reflectivity needed to calculate the absorption rate using a reflection flux monitor based on the total-field scattering-field method. The transmissivity calculation involves placing the flux monitor beneath the ITO ground. Due to the high conductivity and thickness of the ITO ground, the transmissivity consistently equals zero. Therefore, we calculate the absorption rate $A$ using $A = 1 - R - T$, where $R$ represents reflectivity and $T$ represents transmission.

\subsubsection{3.2.1. Optimization with particle swarm optimization (PSO)}

\begin{figure}[!htp]
\includegraphics[width=\textwidth]{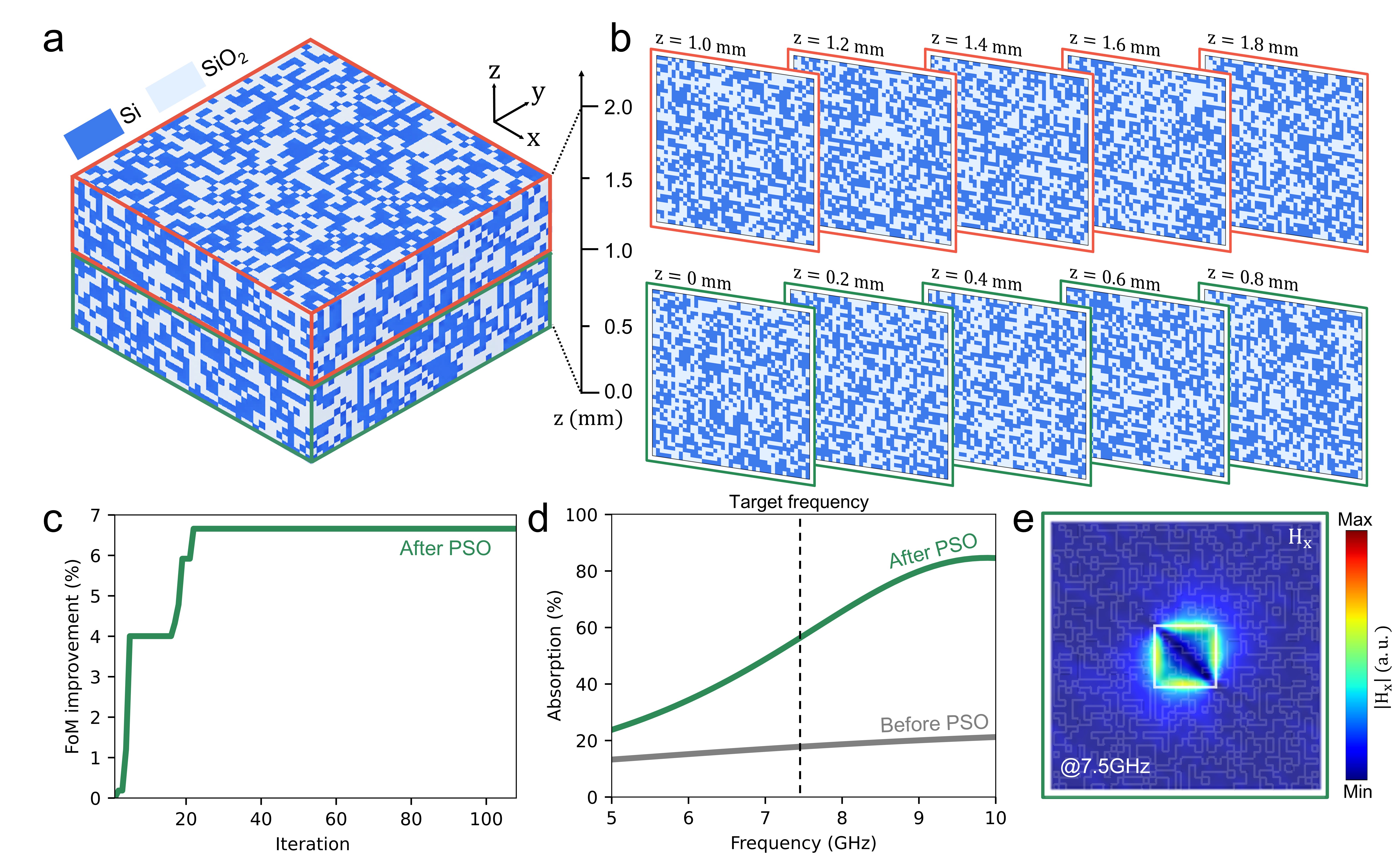}
\caption{%
Layer-by-layer visualization and performance of the 3D PSO design space.
(a) Example of the optimized 3D design domain (5\,mm $\times$ 5\,mm $\times$ 2\,mm). Dark blue represents Si, while light blue indicates SiO$_2$. 
(b) Cross-sectional slices at various heights from $z=0$\,mm to $z=2$\,mm, illustrating the discretized structure with a minimum grid size of 0.125~mm. 
(c) Evolution of the FoM over the optimization iterations. A rapid performance increase is observed compared to the initial structure, followed by convergence to the optimal design. 
(d) Comparison of the absorption spectra before (gray) and after (red) optimization, showing a substantial improvement around 10\,GHz. 
(e) Example of the magnetic field distribution ($|H_z|$) in the final optimized structure. Strong field concentration near the ITO patch leads to high absorption efficiency, even with an ultrathin configuration of $\lambda/20$. 
}
\label{fig6}
\end{figure}

PSO is one of the global optimization methods that explore complex design space with multiple agents~\cite{kennedy1995particle}. However, its computational cost increases dramatically with the number of design variables. To address this limitation, we employed a modified model, binary-Particle Swarm Optimization Gravitational Search Algorithm (binary-PSOGSA)~\cite{mirjalili2014binary}, which integrates the global search capability of PSO with the local refinement performance of the gravitational search algorithm.\\
Fig.~\ref{fig6} summarizes the optimization outcome and electromagnetic characteristics of the 3D absorber design obtained via PSO. The total number of design variables exceeds 30,000, reflecting the high complexity of this 3D configuration. As illustrated in Fig.~\ref{fig6}(a), the final absorber domain measures 5\,mm~$\times$~5\,mm in the $xy$-plane and 2\,mm in the $z$-direction ($\lambda/20$ at 7.5\,GHz). The central ITO patch, with dimensions of 1\,mm~$\times$~1\,mm, is not highlighted in this view but serves as the primary absorption region. Si is represented in dark blue, while SiO$_2$ is depicted in light blue. Fig.~\ref{fig6}(b) presents cross-sectional slices at multiple heights from $z=0$\,mm to $z=2$\,mm, each discretized at a minimum grid size of 0.125\,mm. The binarized distribution of Si and SiO$_2$ appears randomly distributed, indicating that PSO converged to a local optimum with relatively irregular features. In Fig.~\ref{fig6}(c), FoM is plotted against the optimization iterations. Starting from a baseline value of 1.62, the FoM rapidly increases and ultimately converges to 1.73 by the 22\textsuperscript{nd} iteration. This corresponds to a maximum improvement of 6.66\% during the optimization, with the final improvement relative to the initial design also registering at approximately 6.66\%. Fig.~\ref{fig6}(d) compares the absorption spectra of the initial structure (gray) with the PSO-optimized design (green). At the target frequency of 7.5~GHz, the initial structure achieves only 17.80\% absorption, whereas the optimized structure attains 56.28\%, marking a substantial performance gain—an improvement of approximately 3.16$\times$ compared to the initial design. Fig.~\ref{fig6}(e) shows the $|H_z|$ field distribution in the $xy$-plane at mid-height of the absorber. A pronounced field concentration emerges around the edges of the 1\,mm~$\times$~1\,mm ITO patch, indicating that localized resonance significantly enhances absorption.

\subsubsection{3.2.2. Optimization with the adjoint method}

\begin{figure}[!htp]
\includegraphics[width=\textwidth]{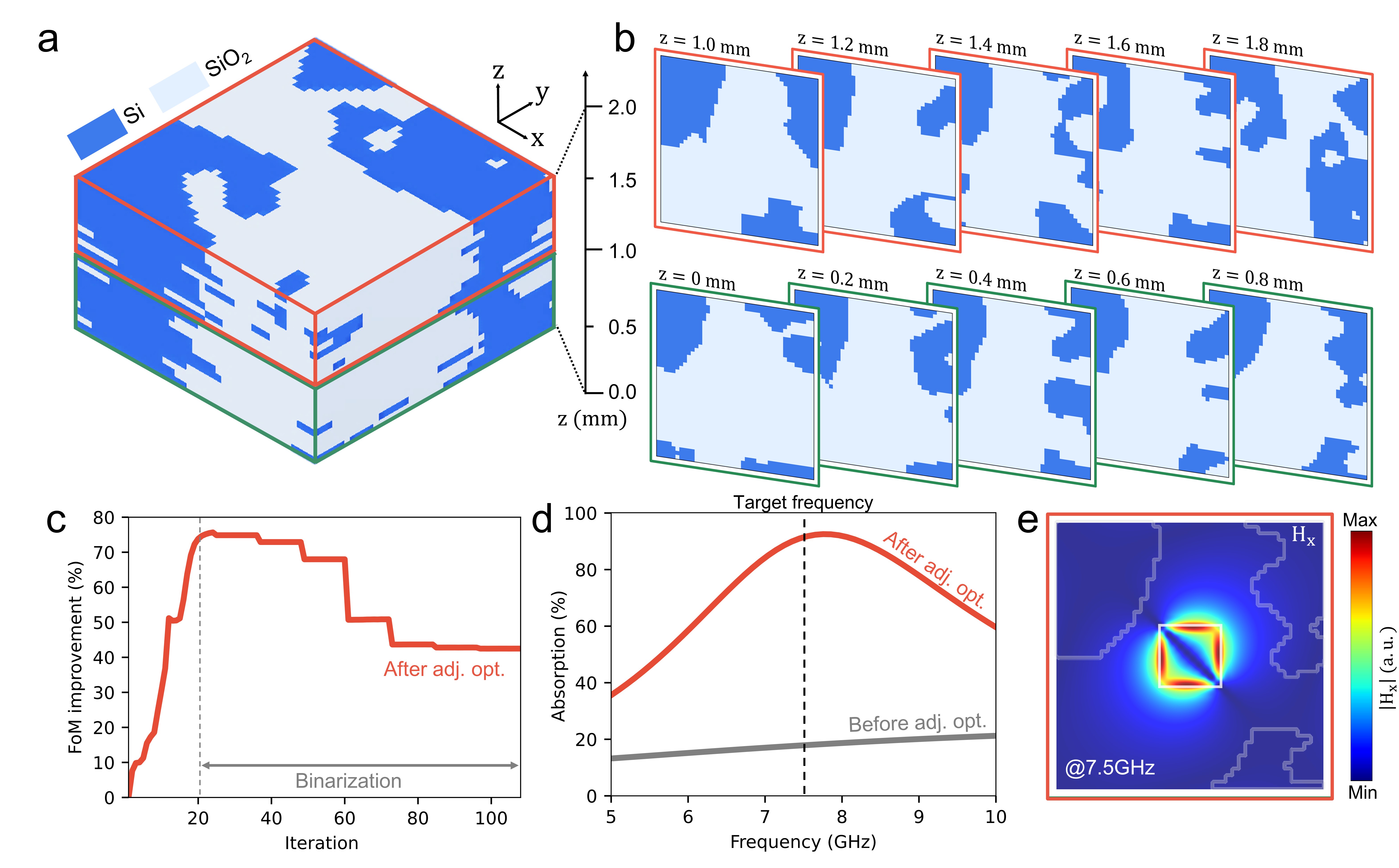}
\caption{Layer-by-layer visualization and performance of the 3D adjoint optimization design space. (a) Example of the optimized 3D design domain (5\,mm $\times$ 5\,mm $\times$ 2\,mm). Dark blue represents Si, while light blue indicates SiO$_2$. (b) Cross-sectional slices at various heights from $z=0$\,mm to $z=2$\,mm, taken at 0.1\,mm intervals, illustrating the discretized structure with a minimum grid size of 0.125\,mm. (c) Evolution of the FoM over the optimization iterations. The FoM rises sharply from the initial structure and is subsequently refined during the binarization process. (d) Comparison of the absorption spectra before (gray) and after (green) optimization, revealing a marked improvement around 7.6\,GHz.
(e) Example of the magnetic field distribution ($|H_z|$) in the final optimized structure. Strong field concentration near the ITO patch enables high absorption efficiency, despite the ultrathin design.
 }
\label{fig7}
\end{figure} 

Using the same design environment employed for the PSO-based optimization, we applied adjoint optimization to design a 3D metamaterial absorber. As described in the “2. Methods” section, adjoint optimization leverages Lorentz reciprocity and the Born approximation for efficient gradient computation. Requiring only two simulations per iteration to calculate the gradient over the design domain, this method optimizes the structure with significantly lower computational cost compared to conventional approaches.
As demonstrated in Fig.~\ref{fig7}, the optimized design obtained via the adjoint method exhibits a well-defined pattern (Fig.~\ref{fig7}(a)), in contrast to the disordered random pattern produced by PSO. This improvement is attributed to the gradient-based nature of adjoint optimization, where the gradient profile over the 3d region changes gradually as it is governed by wave equations. The gradual profile of the gradient directly quantifies the impact of small changes in design variables and steers the design toward an optimal configuration with relatively continuous material distribution.\\ 
Fig.~\ref{fig7}(b) presents cross-sectional slices at multiple heights from $z=0$\,mm to $z=2$\,mm, each discretized at a minimum grid size of 0.125\,mm. Unlike the irregular patterns observed in the PSO-optimized design, the final binarized distribution of Si and SiO$_2$ here appears relatively continuous, indicating that the adjoint method converged to a well-defined configuration with minimal randomness. This outcome highlights how the gradient-based nature of adjoint optimization systematically refines the design while controlling the material layout. In Fig.~\ref{fig7}(c), the FoM is plotted against the optimization iterations. Starting from a baseline value, the FoM initially exhibits a maximum improvement of 75.67\% during the early iterations. Subsequently, after the binarization process is introduced to enforce discrete Si or SiO$_2$ assignments, the final adjoint-optimized design converges to a state reflecting a 42.49\% improvement over the initial configuration.\\
This behavior underscores the efficiency of adjoint optimization in rapidly steering the design toward an optimal solution, while also indicating that the binarization process imposes constraints that stabilize the performance. The robust and stable convergence of the FoM demonstrates the effectiveness of the adjoint approach.\\
Fig.~\ref{fig7}(d) compares the absorption performance before and after the adjoint optimization. The initial structure exhibited an absorption rate of approximately 17.80\% at the target frequency of 7.5\,GHz, indicating that most of the incident electromagnetic energy is reflected rather than absorbed. After applying adjoint optimization, the optimized structure shows that the absorption rate significantly increased to approximately 91.00\% at 7.5\,GHz where it is a relative improvement of $5.11~\times$ compared to the initial design. 

Furthermore, peak absorption improved dramatically from 21.23\% (initial structure) to 92.54\% (optimized structure), corresponding to a 335.9\% enhancement. The absorption peak frequency for the optimized structure is observed at 7.7\,GHz, closely matching the target frequency. Additionally, the bandwidth over which absorption exceeded 90\% ranged from approximately 7.4\,GHz to 8.2\,GHz (a bandwidth of 0.8\,GHz), and the frequency band maintaining absorption above 80\% extended from approximately 6.8\,GHz to 8.9\,GHz (2.1\,GHz bandwidth). 

Although this work focuses on the optimization of the Salisbury screen-integrated MMA for a single frequency, it can be simply expanded to broadband absorber designs because the adjoint optimization offers multi-objective optimizations~\cite{chung2020high}.
Significant differences were also observed in the electromagnetic field distributions. As illustrated in Fig.~\ref{fig7}(e), the structure optimized via the adjoint method exhibited a distinct concentration of the magnetic field localized around the ITO patch, implying that strong surface plasmon resonance (SPR) may occur at the surface area of the ITO. In contrast, the absorber structure designed by PSO displayed a relatively smooth magnetic field distribution around the patch edges, implying weaker SPR effects. These findings highlight not only the computational efficiency but also the superior optimization performance of the adjoint method in metamaterial absorber design. 

Moreover, although electromagnetic absorption is generally proportional to the square of the electric field intensity, the magnetic dipole-based adjoint optimization approach—despite its non-intuitive basis—proves to be a robust and effective alternative, particularly for high-conductivity materials.

\subsubsection{3.2.3. Performance comparison of PSO and adjoint optimization}

\begin{figure}[!htp]
\includegraphics[width=\textwidth]{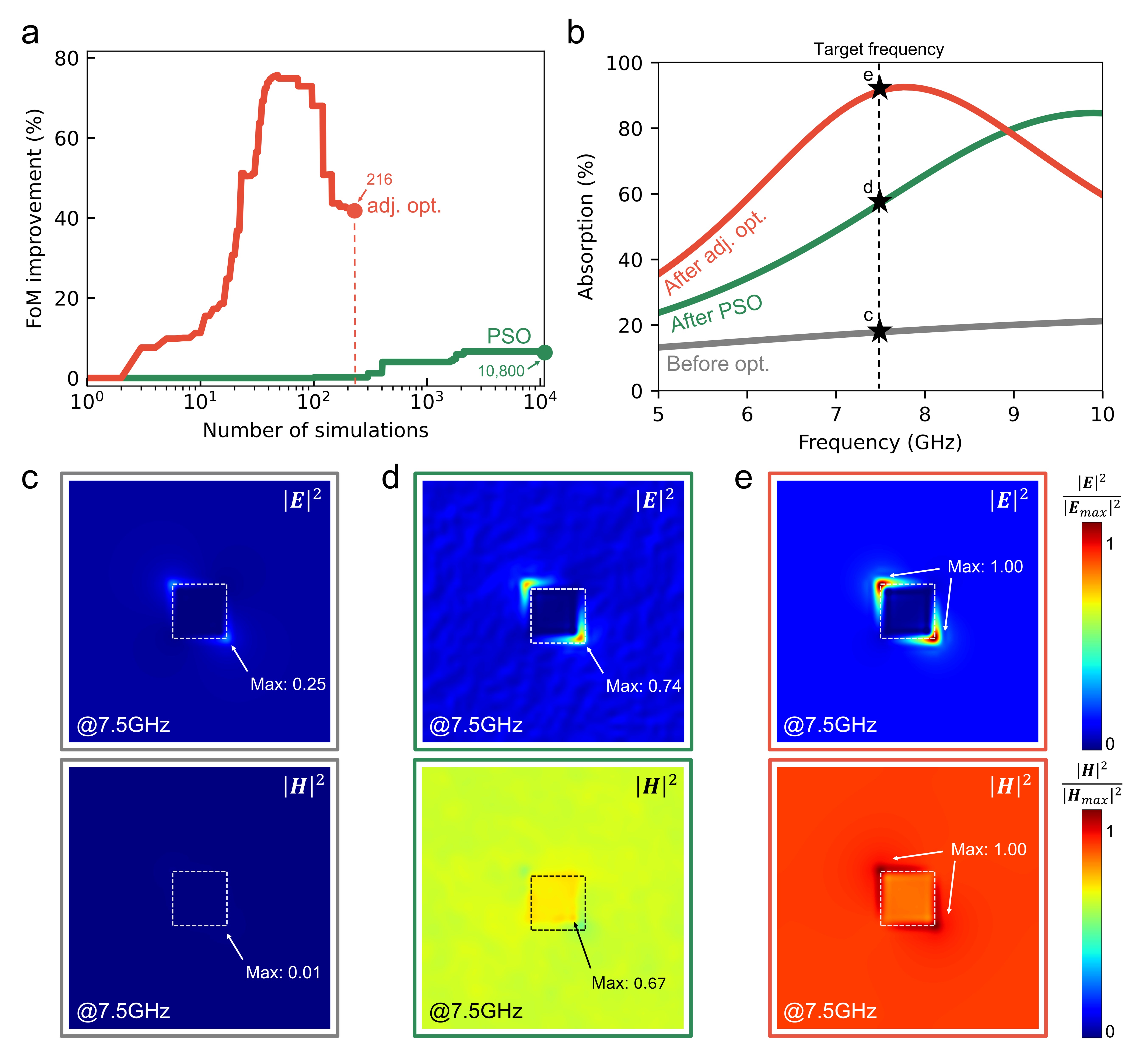}
\caption{
Comparison between PSO and adjoint optimization-based 3D absorber designs.
(a) Evolution of the FoM as a function of simulation count. While the PSO method converges rapidly, its performance gain saturates early. In contrast, adjoint optimization achieves significantly greater improvement with only two simulations per iteration (forward and adjoint).
(b) Absorption spectra of the initial (gray), PSO-optimized (red), and adjoint-optimized (green) structures. The adjoint optimization-based design exceeds 90\% absorption near the target frequency of 7.5~GHz, while the PSO-based design peaks around 9.9~GHz with lower absorption.
(c–e) Normalized electric and magnetic field intensity distributions (\(|E|^2\), \(|H|^2\)) for the three configurations. Intensities are calculated as the sum of squared field components (e.g., \(|H|^2 = |H_x|^2 + |H_y|^2 + |H_z|^2\)) and are shown in arbitrary units (a.u.), normalized to each case’s maximum for visual clarity. The adjoint-optimized structure exhibits strong, localized field enhancement near the ITO patch, enabling efficient absorption in an ultrathin form factor.
}
\label{fig8}
\end{figure} 

This section presents a quantitative comparison between PSO and adjoint optimization for the inverse design of a 3D ultrathin absorber. Performance metrics, including FoM improvement and computational efficiency, are systematically examined. Both optimization methods are assessed under identical simulation conditions, targeting a frequency of 7.5 GHz with an absorber thickness of \(\lambda/20\). More detailed information on the FDTD simulation setup can be found in the Methods section.

Fig.~\ref{fig8}(a) illustrates the progression of FoM as a function of simulation count for each optimization method. While we use 100 particles in PSO over 108 iterations, resulting in a substantial computational cost of 10,800 simulations, adjoint optimization requires significantly fewer simulations—only 216, comprising one forward and one adjoint simulation per iteration. Consequently, adjoint optimization demonstrates a nearly 50-fold computational efficiency advantage over PSO. The absorption spectra for the initial and optimized structures are compared in Fig.~\ref{fig8}(b). At the target frequency of 7.5\,GHz, the initial structure (gray) exhibits an absorption rate of 17.80\%. After optimization, the PSO-designed absorber (green) achieves 56.28\% absorption at 7.5\,GHz, whereas the adjoint-optimized absorber (red) attains a significantly higher absorption rate of 91.00\% at the same frequency.\\
This result clearly demonstrates that, when evaluated at the target frequency, the adjoint-based optimization method yields superior performance compared to PSO, even under stringent $\lambda/20$ thickness constraints.\\
Fig.~\ref{fig8}(c–e) displays the electric (\(E\)) and magnetic (\(H\)) field intensity distributions for the three configurations: initial, PSO-optimized, and adjoint-optimized. The intensity values are computed as \(|E|^2 = |E_x|^2 + |E_y|^2 + |E_z|^2\) and \(|H|^2 = |H_x|^2 + |H_y|^2 + |H_z|^2\), where each component is obtained from the complex field and its conjugate. All values are shown in arbitrary units (a.u.), normalized to the maximum intensity in each case for clarity. Specifically, the maximum \(E\) and \(H\) intensities in the initial geometry are 0.25 and 0.01, respectively; in the PSO-optimized design, they increase to 0.74 and 0.67; and in the adjoint-optimized design, both are normalized to 1. The initial structure exhibits a relatively uniform field distribution across the ITO patch, resulting in weak surface plasmon resonance (SPR) and limited absorption. 

The PSO-based design shows moderate localization near patch edges but remains spatially disordered. In contrast, the adjoint-optimized configuration yields a well-defined and concentrated field distribution, effectively activating SPR and enhancing absorption efficiency even in an ultrathin structure. These results highlight the superiority of adjoint optimization in both computational efficiency and electromagnetic performance.

\section{4. Conclusion}
In this study, we have presented a inverse design framework for ultrathin metamaterial absorbers by effectively integrating the advantageous characteristics of traditional Salisbury screens and metamaterial structures. Through adjoint-based optimization, we successfully designed and optimized absorbers with thicknesses as thin as $\lambda/20$, significantly surpassing traditional constraints of quarter-wavelength thickness. Our optimized structures achieved exceptional absorption efficiencies exceeding 90\% at the targeted frequency of 7.5~GHz, demonstrating superior performance even under oblique incidence conditions.

Furthermore, a comparative analysis against PSO clearly showcased the computational efficiency and superior optimization capabilities of the adjoint method, achieving roughly a 98\% reduction in computational cost. Unlike PSO, which converged to irregular and potentially difficult-to-fabricate designs, adjoint optimization yielded spatially coherent, manufacturable structures with significantly enhanced absorptive properties.

Our work first demonstrated a \(\lambda/20\)-thick Salisbury screen-integrated metamaterial absorber, which may lead to the realization of a compact and lightweight absorbing system. 
These advantages are particularly significant for various practical applications, including stealth applications~\cite{iwaszczuk2011flexible, hossain2022triple}, wireless communications~\cite{karaaslan2017microwave, mohammed2023simulation}, quantum computing~\cite{holman2024trapping}, and numerous other fields highlighted in the introduction.

\section{Author Contribution Statement}
E.J. performed the original draft preparation, manuscript review and editing, methodology design, software development, visualization, validation, data curation, and project administration. J.C. contributed to the original draft preparation, manuscript review and editing, methodology design, software development, validation, and data curation. C.K. was involved in the original draft preparation, manuscript review and editing, methodology design, software development, visualization, project administration, and supervision. H.C. contributed to the original draft preparation, manuscript review and editing, conceptualization, methodology design, visualization, project administration, supervision, and funding acquisition.

\section{Competing Interests}
The authors declare that they have no known competing financial interests or personal relationships that could have appeared to influence the work reported in this paper.

\section{Data availability}
All Data used in this study will be made available upon reasonable request to the corresponding authors.


\nocite{*}
\bibliography{references}
\end{document}